\documentclass[10pt,usletter]{article}
\usepackage[totalwidth=420pt,totalheight=625pt]{geometry}
\usepackage[english]{babel}
\usepackage{graphicx}
\usepackage{amsmath}
\usepackage[labelsep=period]{caption}
\usepackage[breaklinks=true, colorlinks=true, linkcolor=black, citecolor=black, urlcolor=blue]{hyperref}
\usepackage{xfrac}
\usepackage{booktabs}
\usepackage{siunitx}
\usepackage{tikz}

\title{Unmediated AI-Assisted Scholarly Citations\thanks{This is the
    author's version of a paper that appeared in the proceedings of
    The Second Bridge on Artificial Intelligence for Scholarly
    Communication (AAAI-26), \url{https://doi.org/10.52825/ocp.v8i.3161}}}

 \author{Stefan Szeider\\[4pt]
  \small Algorithms and Complexity Group\\[-3pt]
  \small TU Wien, Vienna, Austria\\[-3pt]
  \small \href{https://www.ac.tuwien.ac.at/people/szeider/}{www.ac.tuwien.ac.at/people/szeider/}
 }
\date{}

\begin{document}
\maketitle

\begin{abstract}
Traditional bibliography databases require users to
  navigate search forms and manually copy citation data. Language
  models offer an alternative: a natural-language interface where researchers write
  text with informal citation fragments, which are automatically
  resolved to proper references. However, language models are not
  reliable for scholarly work as they generate fabricated
  (hallucinated) citations at substantial rates.

  We present an
  architectural approach that combines the natural-language interface
  of LLM chatbots with the accuracy of direct database access,
  implemented through the Model Context Protocol. Our system enables
  language models to search bibliographic databases, perform fuzzy
  matching, and export verified entries, all through conversational
  interaction.

  A key architectural principle bypasses the language
  model during final data export: entries are fetched directly from
  authoritative sources, with timeout protection, to guarantee
  accuracy. We demonstrate this approach with MCP-DBLP, a server
  providing access to the DBLP computer science bibliography. The
  system transforms form-based bibliographic services into
  conversational assistants that maintain scholarly integrity. This
  architecture is adaptable to other bibliographic databases and
  academic data sources.
\end{abstract}


\section{Introduction}\thispagestyle{empty}

\subsection{The Opportunity}

Bibliographic databases provide authoritative metadata for scholarly publications but require users to navigate search interfaces, fill in form fields, and manually copy citation data. Consider a researcher writing a manuscript with informal citations like: %

\smallskip
\sloppypar\noindent\texttt{Transformers have revolutionized NLP (Vaswani et al.~2017), with subsequent work on BERT (Devlin paper from 2018) and GPT architectures (Brown et al.~neurips'20)...}
\smallskip

\noindent This process requires the researcher to open the database website, search for each reference using partial information (``Devlin paper from 2018'', ``neurips'20''), identify the correct publication from the results, and copy the citation data, repeating this for every citation.

Language models offer a compelling alternative: a conversational interface where the researcher provides draft text with citation fragments and receives properly formatted references. The language model understands informal citations, searches the database, handles ambiguities through clarification questions, and produces the final bibliography. This approach transforms bibliographic services from form-based tools into interactive assistants that understand research context.

\subsection{The Obstacle}

Language models generate fabricated citations (``hallucinations''). Agrawal et al.~\cite{Agrawal2024} tested \mbox{GPT-3}, ChatGPT, and GPT-4, finding error rates declining from over 70\% to under 50\% with more advanced models. Kim et al.~\cite{Kim2025} examined citations from four models and found fabrication rates exceeding 80\% in certain publication categories. These fabrications include plausible but nonexistent author names, venues, and DOIs. Simply connecting a language model to a database API does not solve this problem, as the model may still fabricate or corrupt bibliographic data while processing search results. %

Several approaches address citation quality. Agrawal et al.~\cite{Agrawal2024} proposed indirect-query self-consistency checks that reduce false discovery rates by 20--30\%. The ALCE benchmark~\cite{Gao2023} evaluates citation quality in long-form question answering, showing that even the best systems provide complete supporting citations for only about 50\% of answers. Aly et al.~\cite{Aly2024} use weakly-supervised fine-tuning to improve citation F1 scores by 34.1 points. Ye et al.~\cite{Ye2024} introduced AGREE, combining tuning with test-time adaptation. Li et al.~\cite{Li2024} introduce Citation-Enhanced Generation, regenerating unsupported claims until citations back each sentence. Examples of detection methods include RefChecker~\cite{Hu2024} and CiteFix~\cite{Maheshwari2025}. Despite these efforts, language models remain unreliable producers of bibliographic data.

\subsection{The Solution}

We present an architectural approach that preserves the conversational interface while guaranteeing citation accuracy. The key insight is to separate natural-language interaction from data retrieval. The language model is responsible for understanding informal citations, disambiguating references through dialogue, and managing iterative search. Databases provide accurate metadata. By connecting these components through a protocol that keeps authoritative data retrieval outside the language model's generation loop, we achieve both natural interaction and verified accuracy. %

We implement this architecture using the Model Context Protocol (MCP)~\cite{Anthropic2024}, a standardized interface for connecting language models with external tools. The MCP enables stateful interactions in which servers provide tools that language models invoke via structured requests. Since its November 2024 release, over 7,000 MCP servers have been developed~\cite{Meetanshi2025}.

Our approach is based on two architectural principles. First, \emph{conversational search through tool calling}: for each natural-language query, language models invoke search tools with extracted parameters. The tools return structured results that models can present, filter, or use for follow-up searches. Second, \emph{unmediated database export}: BibTeX entries are retrieved directly from the database, with timeout protection, and written to files, bypassing the language model entirely. Only citation key replacement occurs outside the database, namely through deterministic pattern matching. This ensures that author names, titles, venues, and DOIs come directly from authoritative sources. %

We demonstrate this architecture with \emph{MCP-DBLP}, a server providing conversational access to the DBLP computer science bibliography (over 6 million publications). The architectural principles apply to any bibliographic database with a programmatic interface (PubMed, arXiv, Semantic Scholar, or institutional repositories). MCP-DBLP implements eight tools: instructions retrieval, boolean search, fuzzy title and author matching, venue information retrieval, statistical analysis, and a two-step BibTeX export. The results show that MCP-DBLP with unmediated export achieves an 82.7\% perfect-match rate, compared to 28.2\% for standard web search, with zero cases of metadata corruption across 104 obfuscated citations across three experiments. %

The system supports both interactive scholarly writing via AI chat applications (Claude Desktop, Cursor, etc.) and autonomous research agents that require verified citation capabilities. The standardized MCP interface makes it composable with other research tools, enabling multi-agent workflows where citation management integrates with literature analysis, writing assistance, and fact-checking.

MCP-DBLP is available at PyPI (\url{https://pypi.org/project/mcp-dblp/}), and the source code is available at \url{https://github.com/szeider/mcp-dblp}. The repository includes installation instructions, an instruction prompt for LLM guidance, and a test suite with 49 automated tests.

\section{Related Work}

The Model Context Protocol ecosystem contains numerous servers for
research applications. For literature discovery, servers exist for
arXiv, PubMed, Semantic Scholar, and OpenAlex (see Appendix~A for
repository listings). The Scientific-Papers-MCP server aggregates
multiple sources, including arXiv, OpenAlex, PubMed Central, bioRxiv,
and medRxiv. For personal library management, multiple Zotero MCP
servers provide vector similarity search, PDF annotation extraction,
and cloud synchronization. Released in February 2025, MCP-DBLP was among the first bibliography-focused MCP servers. %

Most MCP scholarly servers focus on search and metadata retrieval and therefore do not provide citation export. Among those that do, some provide multi-format export (BibTeX, JSON, CSV) or support multiple citation styles (RIS, BibTeX, APA, MLA). However, these servers return formatted citation text in tool responses, meaning the language model receives and could inadvertently modify the citation data before presenting it to users. In contrast, MCP-DBLP implements \emph{unmediated export}: the export tool writes bibliographic data directly to files on disk and returns only file paths to the language model, bypassing the model's context entirely. This architectural choice eliminates citation corruption and reformatting during the export workflow.

\section{System Architecture}

\subsection{Overview}

MCP-DBLP implements the Model Context Protocol server interface, enabling any MCP-compatible client to access DBLP through eight tools. The language model automatically selects appropriate tools based on user queries; users interact through natural language without choosing tools directly. The protocol operates as a stateful client-server architecture where the server maintains session context across multiple tool calls. This allows language models to search, refine queries, and export results through sequential tool invocations. %

\begin{figure}[h]
\centering
\begin{tikzpicture}[
    box/.style={draw, minimum width=1.8cm, minimum height=0.8cm, align=center},
    every node/.style={font=\small}
]
\node[box] (user) {User};
\node[box, right of=user, xshift=2.5cm] (chatbot) {AI Chatbot};
\node[box, right of=chatbot, xshift=2.5cm] (mcpdblp) {MCP-DBLP};
\node[box, right of=mcpdblp, xshift=2.5cm] (dblp) {DBLP};

\draw[<->, thick] (user) -- node[above] {\footnotesize NL} (chatbot);
\draw[<->, thick] (chatbot) -- node[above] {\footnotesize MCP} (mcpdblp);
\draw[<->, thick] (mcpdblp) -- node[above] {\footnotesize API} (dblp);
\end{tikzpicture}
\caption{System workflow: users interact with an AI chatbot, which communicates with MCP-DBLP via the Model Context Protocol. MCP-DBLP queries DBLP via its public API.}
\label{fig:workflow}
\end{figure}
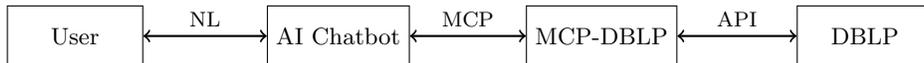

The system is implemented in Python 3.11+ and uses the official MCP SDK. It communicates with clients via standard input/output streams using JSON-RPC 2.0, making it compatible with desktop applications such as Claude Desktop and web-based clients via the Anthropic MCP Connector. Figure~\ref{fig:workflow} shows the system architecture.

\subsection{DBLP API Integration}

The DBLP API provides JSON-formatted publication metadata for each entry, including
titles, authors, venues, years, DOIs, and URLs. MCP-DBLP wraps this
API with error handling, timeout protection (10 seconds for all
requests), and result filtering. If the API returns a timeout or error, the system provides informative error messages. %

\subsection{Tool Implementation}

The server exposes eight tools through the MCP interface (Table~\ref{tab:tools}). Search tools accept query strings, year ranges, venue filters, and similarity thresholds (0.0 to 1.0) for fuzzy matching using Python's \texttt{difflib.SequenceMatcher}. Statistical tools compute aggregate metrics including publication counts, time ranges, top authors, and top venues.

\begin{table}[h]
\centering
\small
\begin{tabular}{@{}llp{4cm}@{}}
\toprule
\textbf{Tool} & \textbf{Function} & \textbf{Key Parameters} \\
\midrule
\texttt{get\_instructions} & Usage guidance and workflow & (none) \\
\texttt{search} & Boolean search with and/or & query, max results, year range, venue filter \\
\texttt{fuzzy\_title\_search} & Fuzzy title matching & title, similarity threshold, filters \\
\texttt{get\_author\_publications} & Author search with fuzzy matching & author name, similarity threshold \\
\texttt{get\_venue\_info} & Venue metadata retrieval & venue name \\
\texttt{calculate\_statistics} & Aggregate statistics & publication list \\
\texttt{add\_bibtex\_entry} & Add entry to collection & dblp\_key, citation\_key \\
\texttt{export\_bibtex} & Write collection to .bib file & path \\
\bottomrule
\end{tabular}
\caption{MCP-DBLP tool specifications}
\label{tab:tools}
\end{table}

\subsection{Unmediated BibTeX Export Mechanism}

The BibTeX export implements a collection-based workflow that bypasses the language model entirely. The workflow is analogous to an online shopping cart: the language model calls \texttt{add\_bibtex\_entry(dblp\_key, citation\_key)} immediately after finding each publication, adding it to a session-scoped collection, and later calls \texttt{export\_bibtex()} to write all collected entries to a file. This ``add to cart'' pattern ensures that the DBLP key is still fresh in the model's context, reducing the risk of key corruption that would occur if the model had to recall keys from earlier in the conversation. %

For each \texttt{add\_bibtex\_entry} call, the server constructs the DBLP URL (\texttt{https://dblp.org/rec/\{dblp\_key\}.bib}) and fetches the BibTeX entry with a 10-second timeout. It replaces the citation key using a regular expression and adds the entry to the collection. This design provides immediate verification and allows the language model to react to errors. If the DBLP key is invalid or the fetch fails, an error is reported back to the language model, which can then retry or inform the user. In the ``checkout'' step, \texttt{export\_bibtex(path)} writes the collection to the specified file path and returns a reference to the resulting file. The bibliographic data (author names with special characters, titles, venues, DOIs) comes directly from DBLP without language model interpretation. An instruction prompt explaining tool usage and best practices is available through the MCP prompt interface.

\section{Use Cases}

The primary use case involves researchers writing papers with AI assistance through applications like Claude Desktop. A researcher can instruct the language model to: (1) identify informal citations in a draft; (2) search DBLP for each reference using fuzzy matching; (3) generate proper citation keys; and (4) export BibTeX entries. The language model uses MCP-DBLP's search and fuzzy\_title\_search tools to find matching publications, presents candidates for user confirmation, and then calls export\_bibtex with the chosen citation keys. This gives a verified .bib file ready for LaTeX compilation. %

Suppose a researcher already has citations in BibTeX, accumulated over time with non-uniform formatting: some conference names are abbreviated, some are in full, some include date and location, some do not. The language model enables the researcher to replace the .bib file with a fresh, certified-correct version in a uniform format. This workflow motivates the use of stateful MCP interactions. In particular, the language model can iterate through multiple searches, handle ambiguous matches through conversation with the user, and finally export only the confirmed citations.

MCP-DBLP also supports autonomous agents that need citation capabilities. For instance, a \emph{deep research} agent tasked with ``write a survey of recent work on graph neural networks'' can use MCP-DBLP to search for relevant papers, extract key contributions through web search or paper access tools, and generate a properly cited survey with verified references. The MCP protocol's tool-description system and the server's instruction prompt provide the necessary context for language models to use DBLP tools correctly in autonomous workflows.

\section{Evaluation}

\subsection{Methodology}

We evaluated bibliography retrieval with and without MCP-DBLP access on 104 obfuscated academic citations across three independent experiments~\cite{Szeider2025Zenodo}. The baseline (\emph{Web}) used only web search, while MCP-enabled methods used DBLP search with either manual BibTeX construction (\emph{MCP-M}) or direct unmediated export (\emph{MCP-U}). %

Ground truth consisted of 104 papers sampled from DBLP, with stratified sampling: 50\% from 2020--2025, 25\% from 2015--2019, and 25\% from 2010--2014. All BibTeX entries were fetched directly from DBLP with a 100\% success rate. Input citations were obfuscated at varying difficulty levels. Difficulty ranged from full author names with years to minimal topic hints. Example inputs include ``Grassi's paper on computer virus from 2025'' and ``hybrid algorithm paper on auvs task by Sun 2024.''

For each experiment, we used Claude Code subagents (Claude Sonnet 4.5) with three configurations: Web (no MCP-DBLP), MCP-M (mediated), and MCP-U (unmediated). The Web baseline relied solely on the WebSearch and WebFetch tools. MCP-M used the MCP-DBLP search tools but manually constructed BibTeX entries, passing citation data through the language model context. MCP-U used the collection-based export API, which writes bibliographic data directly to files.

Citations were classified using a 6-category framework: \emph{Perfect Match} (PM) indicates correct retrieval with all core fields correct; \emph{Wrong Paper} (WP) indicates a different paper than ground truth; \emph{Not Found} (NF) indicates missing or failed retrieval; \emph{Incomplete Metadata} (IM) indicates missing doi, pages, volume, or abbreviated names; \emph{Incomplete Author} (IA) indicates truncated author list; \emph{Corrupted Metadata} (CM) indicates wrong values in fields.

\subsection{Results and Discussion}

Table~\ref{tab:results} shows averaged results across the three experiments. Our results show that MCP-DBLP eliminates metadata corruption entirely. Both MCP-M and MCP-U achieved CM = 0\%, compared to 6.7\% for Web. The CM errors in Web are not typos but plausible fabrications. The LLM ``corrects'' unfamiliar names to common variants (e.g., ``Ma'mon Abu Hammad'' becomes ``Manal Abu Hammad'') or invents page numbers when actual values are unavailable. %

\begin{table}[h]
\centering
\small
\begin{tabular}{@{}lS[table-format=2.1,table-space-text-post=\%]S[table-format=2.1,table-space-text-post=\%]S[table-format=2.1,table-space-text-post=\%]l@{}}
\toprule
\textbf{Category} & {\textbf{Web}} & {\textbf{MCP-M}} & {\textbf{MCP-U}} & \textbf{Description} \\
\midrule
PM & 28.2\% & 47.1\% & 82.7\% & Perfect Match \\
WP & 18.6\% & 15.1\% & 15.7\% & Wrong paper (ambiguous query) \\
NF & 30.1\% & 1.3\% & 1.6\% & Not found \\
IM & 11.9\% & 36.5\% & 0.0\% & Incomplete metadata \\
IA & 4.5\% & 0.0\% & 0.0\% & Incomplete authors \\
CM & 6.7\% & 0.0\% & 0.0\% & Corrupted metadata \\
\bottomrule
\end{tabular}
\caption{Averaged results across three experiments (104 citations each). Web = web search only, MCP-M = MCP-DBLP with mediated export, MCP-U = MCP-DBLP with unmediated export.}
\label{tab:results}
\end{table}

MCP-DBLP also reduces retrieval failures. NF dropped from 30.1\% (Web) to 1--2\% (MCP-M, MCP-U). The remaining NF errors can be attributed to overly vague input citations. In contrast, WP remained at 15--19\% across all methods. This indicates that these errors are due to citation ambiguity rather than the retrieval method. If ``Chaki ieee25'' matches multiple DBLP papers, any method may return an unintended paper.

The comparison between MCP-M and MCP-U reveals a trade-off. MCP-M shows 36.5\% IM because the agent constructs BibTeX entries manually, often omitting DOI, volume, or page numbers. MCP-U achieves 0\% IM by exporting directly from DBLP. This validates unmediated export: every MCP-U entry contains exactly the metadata provided by DBLP.

Our experiments used non-interactive mode, where the agent processed citations autonomously without user feedback. In real-world applications, users would typically ask the language model for clarification when encountering ambiguous references. For instance, when ``Chaki ieee25'' matches multiple DBLP papers, the candidates can be presented to the user to select the intended one. Interactive disambiguation would reduce both WP and NF rates, as the user can provide additional context or confirm matches that the agent would otherwise guess or skip.

\section{Conclusion}

We present an architectural approach for connecting language models with bibliographic databases that combines conversational interaction with verified accuracy. The key insight is separating natural language understanding, handled by language models, from data retrieval, handled by databases. MCP-DBLP implements this approach through the Model Context Protocol, providing conversational access to DBLP with unmediated BibTeX export. %

Our experiments show 82.7\% PM for MCP-U versus 28.2\% for Web, a 2.9$\times$ improvement, with zero metadata corruption, across three independent experiments with 104 obfuscated citations each. The results show that specialized database tools with architectural safeguards can provide both natural language interaction and publication-quality citations. The MCP architecture provides standardization, statefulness, and composability, making the approach applicable across bibliographic databases and research disciplines.




\section*{Acknowledgements}
The author thanks the DBLP team for maintaining the comprehensive
computer science bibliography and providing free API access. The
research was conducted within Cluster of Excellence \emph{Bilateral
Artificial Intelligence} of the Austrian Science Fund (FWF, 10.55776/COE12).

\end{document}